\documentstyle[preprint,aps,epsfig]{revtex}

\begin{document}
\draft
\title{Resonance in the seesaw mechanism}
\author{Haijun Pan$^1$\thanks{Email: phj@mail.ustc.edu.cn} and 
	G. Cheng$^{2,3}$\thanks{Email: gcheng@ustc.edu.cn}}
\address{\it
$^1$Lab of Quantum Communication and Quantum Computation, and Center of
Nonlinear Science,\\
University of Science and Technology of China, Heifei, Anhui, 230026, P.R.China\\
$^2$CCAST(World Laboratory), P.O.Box 8730, Beijing 10080, P.R.China\\
$^3$Department of Astronomy and Applied Physics, \\
University of Science and Technology of China, Heifei, Anhui, 230026, P.R.China}
\date{\today}
\maketitle

\begin{abstract}
We study the RH neutrino properties from the low energy neutrino data in the 
seesaw mechanism. Reonance behavior is found for the right-handed (RH) mixing 
angle as a function of light neutrino mass ratios in two favored region of the 
solar neutrino problem and, at the resonance point, the two corresponding RH 
Majorana neutrino masses are degenerate. This phenomenon is similar with that 
in the matter-enhanced conversion of neutrinos. The physical significance it 
infers for the electron neutrino mass and other neutrino parameters is discussed.
\end{abstract}



\newpage \baselineskip20pt

The smallness of left-handed (LH) neutrino masses implied by the solar and 
atmospheric neutrino experiments \cite{solar,atm} is perhaps attributed to 
the Majorana feature of neutrino fields \cite{bilenky,fritzsch}. By adding 
the right-handed (RH) fields $\nu _{R}$ of heavy neutrinos, the seesaw 
mechanism \cite{ss}\ provides a very natural and attractive explanation of 
the smallness of the neutrino masses compared to the masses of the charged 
fermions. In this mechanism the LH Majorana neutrino mass matrix, $m_{\nu }$, 
is expressed in the following form \cite{smirnov}: 
\begin{equation}
	m_{\nu }=m_{D}M^{-1}m_{D}^{T},  \label{seesaw}
\end{equation}
where $m_{D}$ is the Dirac mass matrix which, as suggested by Grand Unified 
Theories (GUTs), is assumed to be similar to that in the quark sector and 
$M$ is the Majorana mass matrix for the RH neutrino components. After the 
pioneer work of Smirnov, interest in determining the properties of the RH 
Majorana neutrinos has been revised with expanding experimental data about 
neutrinos (see \cite{falcone,kuo,pan1}). 

The Dirac matrix can be diagonalized by the bi-unitary transformation 
\cite{fritzsch}: $m_{D}^{\rm diag}=D_{L}^{\dag }m_{D}D_{R}={\rm diag}\{ 
m_{1D},m_{2D},m_{3D}\} $. $m_{\nu }$ and $M$ are in general complex matrices 
and, respectively, only a single unitary matrix is needed for the diagonalization: 
$m_{\nu }^{\rm diag}=U^{T}m_{\nu }U={\rm diag }\{ m_1,m_2,m_3\} $, 
$M^{\rm diag}=V_{0}^{T}MV_{0}={\rm diag }\{ M_1,M_2,M_3\} $.
Eq.\ (\ref{seesaw}) can be rewritten as 
\begin{equation}
	V(M^{\rm diag})^{-1}V^{T}
	=(m_{D}^{\rm diag})(S^{\dag })^{T}m_{\nu }^{\rm diag}S^{\dag }(m_{D}^{\rm diag})
	\equiv X,
\label{X}
\end{equation}
where $V=D_{R}^{\dag }R$ is the RH mixing matrix containing the contributions 
from the diagonalizations of both $m_{D}$ and $M^{-1}$ and can be 
parameterized by $V=e^{i\beta _{23}\lambda _{7}}e^{i\beta _{13}\lambda _{5}}
e^{i\beta _{12}\lambda _{2}}$ with $\lambda _{2,5,7}$ the Gell-Mann matrices 
(for example, see \cite{para}). $S=D_{L}^{T}U$ is so-called seesaw matrix 
which specifies the feature of the seesaw mechanism \cite{smirnov}. In this 
letter, assuming quark-lepton symmetry and hierarchical Dirac and LH Majorana 
neutrino spectra, we aim at analyzing the RH neutrino properties (masses and 
mixing) from the low energy neutrino data. The CP-violating effect will be 
ignored in our analysis and so all the transformation matrices entered the 
seesaw mechanism will be real orthogonal. Due to the large difference among 
different eigenvalues of $X$ (usually they can be apart away more than about 
$10$ magnitudes), it is hard to obtain them precisely even numerically. Even 
this is done, the complicated expressions of them make it impossible to see 
explicit dependence on various physical parameters. We notice that the left 
side of Eq.\ (\ref{X}) contains the contributions mainly from $M_{1}^{-1}$ 
and $M_{2}^{-1}$ while its inverse $V(M^{\rm diag})V^{T}$ contains the 
contributions mainly from $M_{2}$ and $M_{3}$. Two of the eigenvalues can be 
obtained through solving a quadratic equation and the third is then obtained 
by the equation of the determinants of both sides of Eq.\ (\ref{X})  \cite{pan1}. 

As suggested by Smirnov, $D_{L}$ is nearly a unit matrix and then $U\approx S$ 
which can be parameterized as 
\begin{equation}
U=e^{i\theta _{23}^{\nu }\lambda _{7}}e^{i\theta _{13}^{\nu }\lambda _{5}}
	e^{i\theta _{12}^{\nu }\lambda _{2}}.
\end{equation}
Among the three angles, $\theta _{13}^{\nu }$ is small implied by the CHOOZ 
observation  \cite{chooz}, $\theta _{23}^{\nu }$ is almost maximal (i.e. 
$\theta _{23}^{\nu }\approx \frac{\pi }{4}$) which is suggested strongly by 
the recent result from SuperKamiokande \cite{atm} while $\theta _{12}^{\nu }$ 
which is responsible for the solar neutrino deficit can vary from rather small 
value ($10^{-3}$) (for SMA: small angle MSW \cite{msw} effect) to nearly $\pi /4$ 
(for VO: vacuum oscillation, LAM: large angle mixing MSW effect and LOW: low 
mass or possibility)\cite{bahcall}.

When $\theta _{12}^{\nu }$ is small, we consider a special case: when 
$\theta _{12}^{\nu }\sim \theta _{13}^{\nu }$. For convenient, we set $U_{\tau 1}=0$. 
The three RH Majorana masses are given by  \cite{details}
\begin{mathletters}
\begin{eqnarray}
	M_{1} &\approx &\frac{m_{1D}^{2}}{m_{3}}\cot ^{2}\theta _{12}^{\nu }, \\
	M_{2} &\approx 
		&\left\{ 
		\begin{array}{ccc}
			2\frac{m_{2D}^{2}}{m_{1}}\sin ^{2}\theta _{12}^{\nu }, 
			&  & {\rm if\quad }r_{21}<r_{21}^{{\rm res}} \\ 
			\frac{1}{2}\frac{m_{3D}^{2}}{m_{2}}, 
			&  & {\rm if\quad }r_{21}>r_{21}^{{\rm res}}
		\end{array}
		\right. \\
	M_{3} &\approx 
		&\left\{ 
		\begin{array}{ccc}
			\frac{1}{2}\frac{m_{3D}^{2}}{m_{2}}, 
			&  & {\rm if\quad }r_{21}<r_{21}^{{\rm res}} \\ 
			2\frac{m_{2D}^{2}}{m_{1}}\sin ^{2}\theta _{12}^{\nu }, 
			&  & {\rm if\quad }r_{21}>r_{21}^{{\rm res}}
		\end{array}
		\right.
\end{eqnarray}
\label{mass1}
\end{mathletters}
where $r_{21}=\frac{m_{2}}{m_{1}}$ and 
$r_{21}^{\rm res}=\frac{1}{4}\frac{m_{3D}^{2}}{m_{2D}^{2}}\csc ^{2}\theta _{12}^{\nu }$. 
Note that we have two degenerate masses $M_{2}=M_{3}$ 
when $r_{21}=r_{21}^{{\rm res}}$.

The second RH mixing angle $\beta _{23}$ is given in 
\begin{equation}
	\sin 2\beta _{23}
	\approx -\frac{2m_{2D}/m_{3D}}
		{\sqrt{\left( 1-r_{21}/r_{21}^{{\rm res}}\right) ^{2}
			 +4m_{2D}^{2}/m_{3D}^{2}}}.
\end{equation}
The other two RH mixing angles are both small and can be expressed in $\beta _{23}$ 
as follows
\begin{equation}
	\beta _{12}
	\approx \frac{1}{\sqrt{2}\sin \theta _{12}^{\nu }}
	\left( -\frac{m_{1D}}{m_{2D}}\cos \beta _{23}
		+\frac{m_{1D}}{m_{3D}}\sin \beta _{23} 
	\right) ,
\end{equation}
\begin{equation}
	\beta _{13}
	\approx \frac{1}{\sqrt{2}\sin \theta _{12}^{\nu }}
	\left( -\frac{m_{1D}}{m_{2D}}\cos \beta _{23}
		+\frac{m_{1D}}{m_{3D}}\sin \beta _{23}
	\right) .
\end{equation}

The behavior of $\sin ^{2}2\beta _{23}$ (see in Fig.\ \ref{fig1}) as a function of 
$r_{21}$ is clearly that of a resonance peaked at $r_{21}=r_{21}^{{\rm res}}$. 
The phenomenon is very like that in the matter-enhanced $\nu _{e}\leftrightarrow 
\nu _{\mu }$ oscillation in the sun except that, in our case, $r_{21}$ plays a part 
of the effective potential $V=2\sqrt{2}G_{f}N_{e}E_{\nu }$. Here $G_{f}$\ is the 
Fermi constant, $N_{e}$ is the electron number density of the matter and $E_{\nu }$ 
is the neutrino energy. We can define the relative resonance width $\delta $ as that 
of $r_{21}/r_{21}^{\rm res}$ around $r _{21}^{{\rm res}}$ for which 
$\sin ^{2}2\beta _{23}$ becomes $\frac{1}{2}$ instead of the maximum value, unity. 
It is given by
\begin{equation}
	\delta \approx 4\frac{m_{2D}}{m_{3D}}
\end{equation}
which is far less than unit. If the two heavier RH Majorana neutrino are degenerate 
and then the corresponding RH mixing angle $\beta _{23}\approx \pi /4$, one can determine 
the electron neutrino mass 
\begin{equation}
	m_{1}	\approx 1.6\times 10^{-10}~{\rm eV}.
\end{equation}
Around this point, $M_{1}\approx 1.4\times 10^{7}~{\rm GeV}$ and
\begin{equation}
	M_{2}\approx M_{3}\approx 2.6\times 10^{15}~{\rm GeV}
\end{equation}
$M_{2}$ ($M_{3}$) remains at $2.6\times 10^{15}~{\rm GeV}$ and $M_{3}$ ($M_{2}$) increases almost linearly with the decrease of $m_{1}$ for $m_{1}<(>) 1.6\times 10^{-10}~{\rm eV}$ 
and so we can also infer the scale of $m_{1}$ from the knowledge of the spectrum of the RH Majorana neutrinos. The above values are obtained by taking $m_{2}\approx \sqrt{\Delta m_{\rm solar}^2}\approx \sqrt{5.4\times 10^{-6}}~{\rm eV}$ and $\sin ^{2}2\theta _{12}^{\nu }\approx \sin ^{2}2\theta _{\rm solar}\approx 6.0\times 10^{-3}$ \cite{bahcall}. We always 
take $m_{3}\approx \sqrt{\Delta m_{\rm atm.}^2}\approx \sqrt{5.9\times 10^{-3}}~{\rm eV}$ \cite{atm} and $m_{D}^{\rm diag}\approx m^{\rm up}(\mu)$ \cite{fusaoka}. Here $\mu =10^{9}
~{\rm GeV}$.

If $\theta _{12}^{\nu }$ is large, we find, when $\sin ^{2}\theta _{13}^{\nu }\ll \frac{1}{2}\frac{m_{1D}^{2}}{m_{2D}^{2}}$, the RH Majorana masses are given by 
\begin{mathletters}
\begin{eqnarray}
	M_{1} &\approx 
		&\left\{ 
		\begin{array}{ccc}
		\frac{m_{1D}^{2}}{m_{2}}\frac{1}{\sin ^{2}\theta _{12}^{\nu }} 
		&  & \frac{m_{3}}{m_{2}}
			<2\frac{m_{2D}^{2}}{m_{1D}^{2}}\sin ^{2}\theta _{12}^{\nu } \\ 
		2\frac{m_{2D}^{2}}{m_{3}} 
		&  & \frac{m_{3}}{m_{2}}
			>2\frac{m_{2D}^{2}}{m_{1D}^{2}}\sin ^{2}\theta _{12}^{\nu }
		\end{array}
		\right.  \\
	M_{2} &\approx 
		&\left\{ 
		\begin{array}{ccc}
		2\frac{m_{2D}^{2}}{m_{3}} 
		&  & \frac{m_{3}}{m_{2}}
			<2\frac{m_{2D}^{2}}{m_{1D}^{2}}\sin ^{2}\theta _{12}^{\nu } \\ 
		\frac{m_{1D}^{2}}{m_{2}}\frac{1}{\sin ^{2}\theta _{12}^{\nu }} 
		&  & \frac{m_{3}}{m_{2}}
			>2\frac{m_{2D}^{2}}{m_{1D}^{2}}\sin ^{2}\theta _{12}^{\nu }
		\end{array}
		\right.  \\
	M_{3} &\approx &\frac{1}{2}\frac{m_{3D}^{2}}{m_{1}}\sin ^{2}\theta _{12}^{\nu }
\end{eqnarray}
\label{mass2}
\end{mathletters}
and the RH mixing angles $\beta _{13}\approx \frac{\sqrt{2}m_{1D}}{m_{3D}}
\cot \theta _{12}^{\nu }$ and	$\beta _{23}\approx -\frac{m_{2D}}{m_{3D}}$.
We have deduce simple expression for $\beta _{13}$. In Fig.\ \ref{fig2} $M_{1}$, 
$M_{2}$ and $\sin ^{2}2\beta _{12}$ near $r_{32}^{\rm res}\approx 2\frac{m_{2D}^{2}}{m_{1D}^{2}}\sin ^{2}\theta _{12}^{\nu }\sim 10^{4}$ are 
numerically plotted taking $m_{3}^{2}=0.1~{\rm eV}^{2}$. The behavior 
$\sin ^{2}2\beta _{12}$ as a functions of $r_{32}$\ is also a resonance 
peaked at $r_{32}^{{\rm res}}$. Taking $10^{-1}~{\rm eV}^2<m_{3}^{2}\approx 
\Delta m_{\rm atm.}^2<10^{-3}~{\rm eV}^2$ in consider, the domain of $m_{2}^{2}$ corresponding 
to $r_{32}^{\rm res}$ is $10^{-11}-10^{-13}$ which lies in the region of the vacuum 
oscillation (VO) solution to the solar neutrino problem. For the large mixing MSW 
effect and low mass MSW effect, one always has $\frac{m_{3}}{m_{2}}<2\frac{m_{2D}^{2}}{m_{1D}^{2}}\sin ^{2}\theta _{12}^{\nu }$ 
and so 
\begin{equation}
	M_{1}\approx \frac{m_{1D}^{2}}{m_{2}}\frac{1}{\sin ^{2}\theta _{12}^{\nu }},\ \ \ 	M_{2}\approx 2\frac{m_{2D}^{2}}{m_{3}},\ \ \ 
	M_{3}\approx \frac{1}{2}\frac{m_{3D}^{2}}{m_{1}}\sin ^{2}\theta _{12}^{\nu };
\end{equation}
\begin{equation}
	\beta _{12}\approx -\frac{1}{\sqrt{2}}\frac{m_{1D}}{m_{2D}}
				\cot \theta _{12}^{\nu },\ \ \ 	
	\beta _{13}\approx \sqrt{2}\frac{m_{1D}}{m_{3D}}
				\cot \theta _{12}^{\nu },\ \ \ 
	\beta _{23}\approx -\frac{m_{2D}}{m_{3D}}.
\end{equation}
For VO, we have the similar tendency of $M_{1,2}$ with the increase of $r_{32}$ 
as that of $M_{2,3}$ with the increase of $r_{21}$ in the SMA case. So the 
allowed region of the mass-squared difference $\Delta m_{23}^{2}$ could be 
confined in a relatively narrow range from the RH Majorana neutrino spectrum. 
Inserting $m_{2}\approx \sqrt{\Delta m_{solar}^2}\approx \sqrt{6.5\times 10^{-11}}
~{\rm eV}$ and $\sin ^{2}2\theta _{12}^{\nu }\approx \sin ^{2}2\theta _{solar}
\approx 0.75$ \cite{bahcall} in Eq.\ (\ref{mass2}), we obtain $M_{1}\approx M_{2}
\approx 3.6\times 10^{9}~{\rm GeV}$ and $M_{3}\approx 1.8\times 10^{17}r_{21}~
{\rm GeV}$ when $r_{32}\approx 4.1\times 10^{4}$. 

To summarize, from the mass spectrum of RH Majorana neutrino, we can infer 
whether the long wave-length vacuum oscillation or the matter-enhanced conversion 
is responsible for the solar neutrino deficit. If the small mixing MSW effect 
is just the case, the electron neutrino mass may be inferred from the structure 
of mass spectrum of the RH Majorana neutrino. On the other hand, if the solar 
neutrino deficit is due to the vacuum oscillation, the structure of such a 
spectrum could constrain the region of the $m_{2,3}$. However, $M_{3}(\gg 1.8
\times 10^{17}~{\rm eV})$ is too large to be believable in VO case. Note that 
$M_{2,3}$ are near the unification scale ($\sim 10^{16}~{\rm GeV}$) \cite{falcone} 
in the supersymmetric case while $M_{1}$ is even far less than the intermediate 
scale ($\sim 10^{9}-10^{13}~{\rm GeV}$) \cite{falcone,inter}. It would be 
interesting if the information about the spectrum of the RH Majorana neutrino 
can be corroborated by other theories such as the supersymmetry theory or the 
string. 

The results are dependent on the precise determination of $U_{e3}$. Such a goal 
is expected to be reached in the undergoing or forthcoming neutrino long baseline 
experiment, registration of the neutrino bursts from the Galactic supernova by 
existing detectors SK and SNO, and the neutrino factories \cite{goal}. 

\acknowledgements

We would like to express our sincere thanks to Professor T.K. Kuo for his warmly 
help in the research progressing. We are grateful to Dr. Du Taijiao for his careful 
examination of the paper especially the numerical results. The authors also thank 
Dr. Du Taijiao and Dr. Tu Tao for useful discussions. This research is supported 
by the National Science Foundation in China grant No.19875047.

\addtolength{\baselineskip}{-.3\baselineskip}

\newpage 
\begin{figure}[tbp]
\caption{The behavior of the $M_2$, $M_3$ and $\protect\beta_{23}$ as a function
of $\protect\frac{m_2}{m_1}$ for the SMA solution to the solar neutrino problem  
taking $U_{\protect\tau 1}=0$. We take $m_{2}^{2}=5.4\times 10^{-6}~
{\rm eV}^{2}$, $m_{3}^{2}=5.9\times 10^{-3}~{\rm eV}^{2}$, $\sin ^{2}2
\protect\theta _{12}^{\protect\nu }=6.0\times 10^{-3}$, $\protect\theta 
_{13}^{\protect\nu }=0.0$ and $\sin ^{2}2\protect\theta _{23}^{\protect\nu 
}=1.0$.}
\label{fig1}
\end{figure}

\begin{figure}[tbp]
\caption{The behavior of $M_{1}$, $M_{2}$ and $\sin ^{2}2\protect\beta _{12}$
as a function of $\protect\frac{m_3}{m_2}$ for the vacuum oscillation solution
to the solar neutrino problem taking $U_{e3}=0$. $\sin ^{2}2\protect
\theta _{12}^{\protect\nu }=0.75$ and $\sin ^{2}2\protect\theta _{23}^{
\protect\nu }=1.0$.}
\label{fig2}
\end{figure}

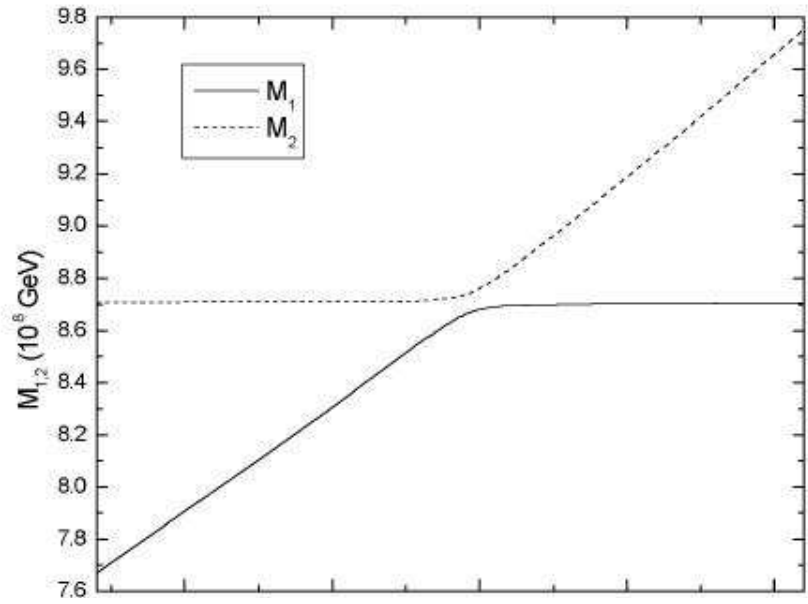
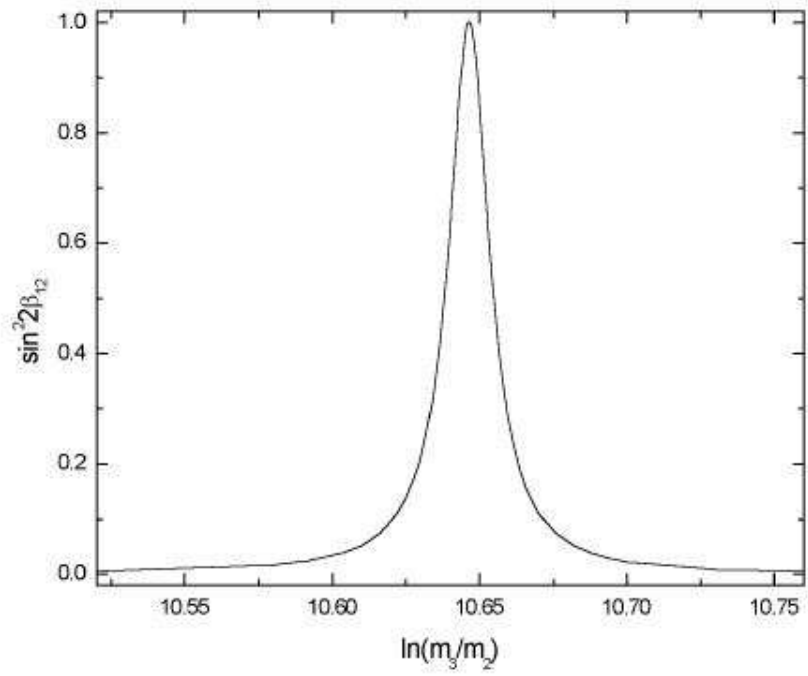
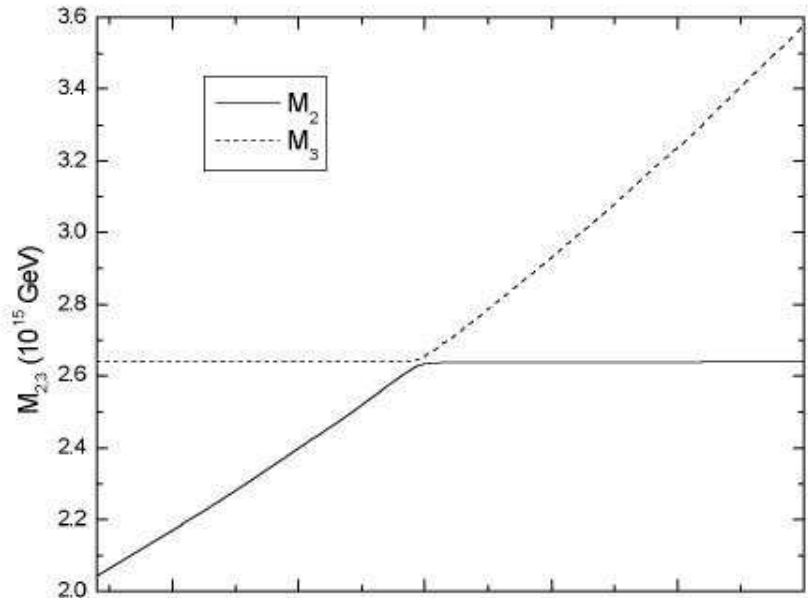
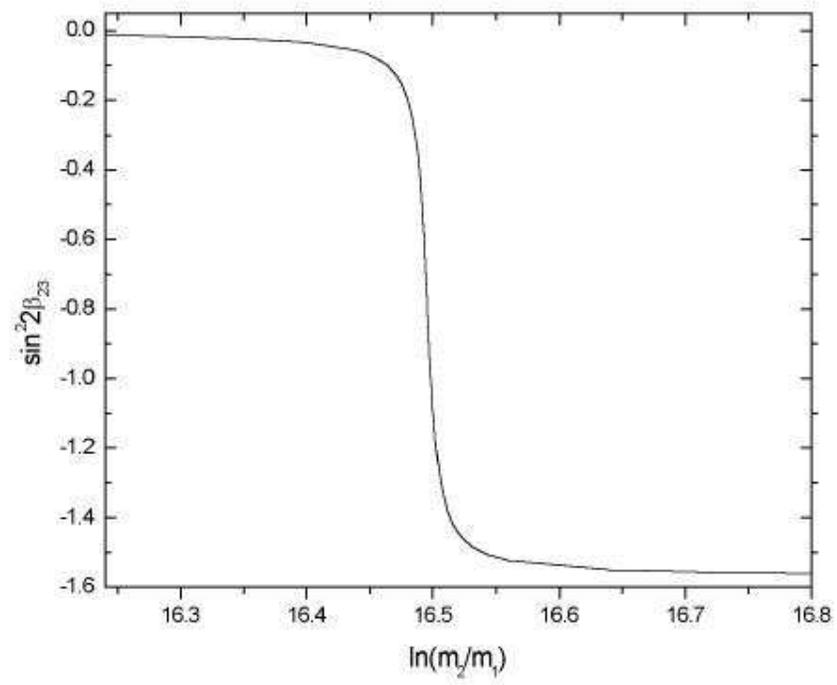
\end{document}